\newcounter{myctr}
\def\myitem{\refstepcounter{myctr}\bibfont\noindent\ifnum\themyctr>9\else\phantom{0}\fi\hangindent17pt\themyctr.\enskip}

\documentclass{ws-ijqi}
\usepackage{hyperref}
\usepackage[super,sort,compress]{cite}

\begin{document}
%%%%%%%%%%%%%%%%%%%%% Publisher's Area please ignore %%%%%%%%%%%%%%
\catchline{}{}{}{}{}
%%%%%%%%%%%%%%%%%%%%%%%%%%%%%%%%%%%%%%%%%%%%%%%%%%%%%%%%%%%%%%%%%%%

\title{Quantum coherent transport in a three-arm beam splitter\\ and a Braess paradox}

\author{E. Zhitlukhina}
%Typeset names in8~pt roman, uppercase. Use the footnote to indicate the
% present or permanent address of the author.}
\address{O.O. Galkin Donetsk Institute for Physics and Engineering, 03028 Kyiv, Ukraine\\
Vasyl' Stus Donetsk National University, 21021 Vinnytsia, Ukraine\\ 
elena\_zhitlukhina@ukr.net}
\author{M. Belogolovskii}
\address{G.V. Kurdyumov Institute for Metal Physics, 03142 Kyiv, Ukraine\\
Vasyl' Stus Donetsk National University, 21021 Vinnytsia, Ukraine}
\author{N. De Leo, M. Fretto, A. Sosso}
\address{National Institute for Metrological Research, 10135 Torino, Italy}
\author{P. Seidel} 
\address{Institut f\"{u}r Festk\"{o}rperphysik, Friedrich-Schiller-Universit\"{a}t Jena, 07743 Jena, Germany}

\maketitle

\begin{history}
% \received{Day Month Year}
% \revised{Day Month Year}
%\accepted{Day Month Year}
%\comby{(xxxxxxxxxx)}
\end{history}

\begin{abstract}
The Braess paradox encountered in classical networks is a counterintuitive phenomenon when the flow in a road network can be impeded by adding a new road or, more generally, the overall net performance can degrade after addition of an extra available choice. In this work, we discuss the possibility of a similar effect in a phase-coherent quantum transport and demonstrate it by example of a simple Y-shaped metallic fork. To reveal the Braess-like partial suppression of the charge flow in such device, it is proposed to transfer two outgoing arms into a superconducting state. We show that the differential conductance-vs-voltage spectrum of the hybrid fork structure varies considerably when the extra link between the two superconducting leads is added and it can serve as an indicator of quantum correlations which manifest themselves in the quantum Braess paradox.
\end{abstract}

\keywords{quantum Braess paradox; Y-shaped metallic fork; charge flow; superconducting outgoing leads.}

%\tableofcontents  % optional
%\markboth{Authors' Names}

\section{Introduction}

In 1968, Dietrich Braess published a paper \cite{braess68} where he showed that the flow in a road network can degrade after addition of a new transport channel. The counterintuitive phenomenon appears when each driver is making the optimal decision about which route is quickest. Braess showed that the attempt to minimize own travel time while ignoring the effect of this decision on other travelers can leave all users worse off than before the new link was introduced. The classical network configuration used to illustrate the Braess paradox consists of start (A) and end (B) nodes with two roads, left (ALB) and right (ARB) connecting them, Fig. 1a. When the two ways are identical, a half of drivers prefer to go along the ALB path whereas others select ARB path, in this case the needed time coincides and is, say, $\tau_1$. Now we assume that two segments AL and LB of the ALB road as well as AR and RB parts of the ARB road are not identical and an extremely fast LR road connecting ALB and ARB paths (a dashed line) is built, Fig. 1b. Braess showed \cite{braess68} that under some conditions the drivers, most probably, are choosing the same ARLB route with the time spent along it $\tau_2>\tau_1$ (later it was shown that whether the Braess paradox does or does not occur depends on the problem parameters, i.e., the effect reveals itself within a certain range of the road characteristics) \cite{pas97}.

This conclusion is considered unexpected since the addition of an extra resource to a network, and therefore an extra available choice, reduces (but not enhances as one might think) the overall performance. In a broader perspective, the explanation of this paradox consists in the fact that the introduction of a new capacity enriches the complexity of the problem. As was shown in Refs. 3 and 4, the dynamics which considers only binary choices is usually so strongly modified after introducing a ternary choice that the final situation cannot be considered as a ``binary choice plus one'' but is much more complicated. 

  The Braess paradox has attracted attention of researchers from different scientific fields, from computer sciences to economics, see Ref. 5 and related references in Ref. 4. There are some macroscopic analogs of this effect in electrical, mechanical, and thermal nets \cite{cohen91,pench03}. The authors of Ref. 8 drew attention to the possibility of realizing a mesoscopic analog of the Braess phenomenon in a semiconductor two-path network with spatial dimensions less than the coherence length which in their devices was of the order of several microns at low temperatures. The question they addressed in the conclusions was whether the phenomenon predicted in Ref. 1 can manifest itself in other coherent systems where transport is governed by quantum mechanics. But, as was emphasized in the paper \cite{sousa13}, before declaring a quantum origin of a phenomenon one must reveal the fundamental reasons behind the transmission reduction phenomena and to show that they are indeed of the quantum origin. The authors of Ref. 9 studied wave packet propagation through a circular quantum ring attached to input and output leads in the presence of an extra channel passing diametrically through the ring and certified that the transport inefficiency in such system originates from the quantum scattering of the wave packet in the input channel-ring junction and the quantum interference between parts of the wave packets that passed through the central channel and those that propagated through the ring arms.  
  
  In our work, we propose a novel quantum analog of the Braess paradox with non-classical correlations. It is strongly reminiscent of the road network in the original paper \cite{braess68} and can be realized in a simple Y-shaped conﬁguration with a normal-metal (N) incoming electrode (an emitter) and two superconducting (S) outgoing leads connected to a collector, see Fig. 1c. 
\begin{figure}[bt]
\centerline{\includegraphics[width=.98\columnwidth]{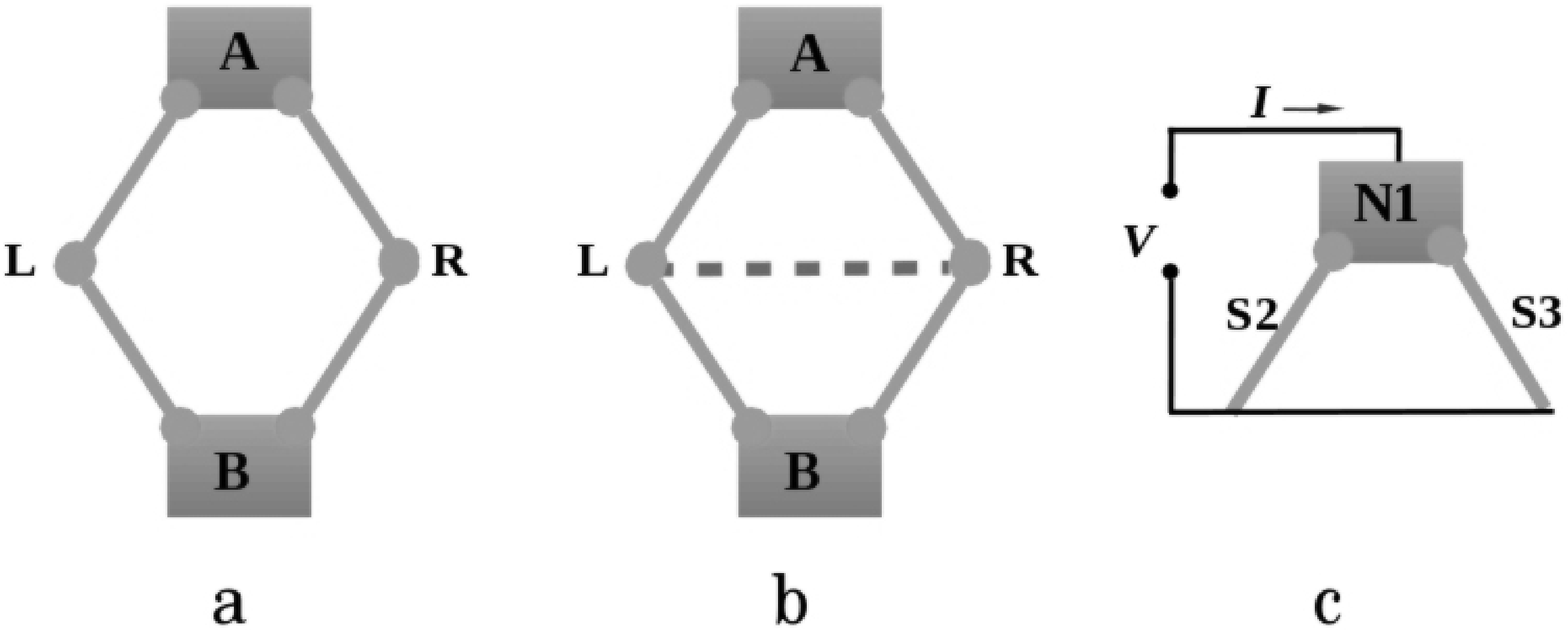}}
\centerline{\includegraphics[width=.98\columnwidth]{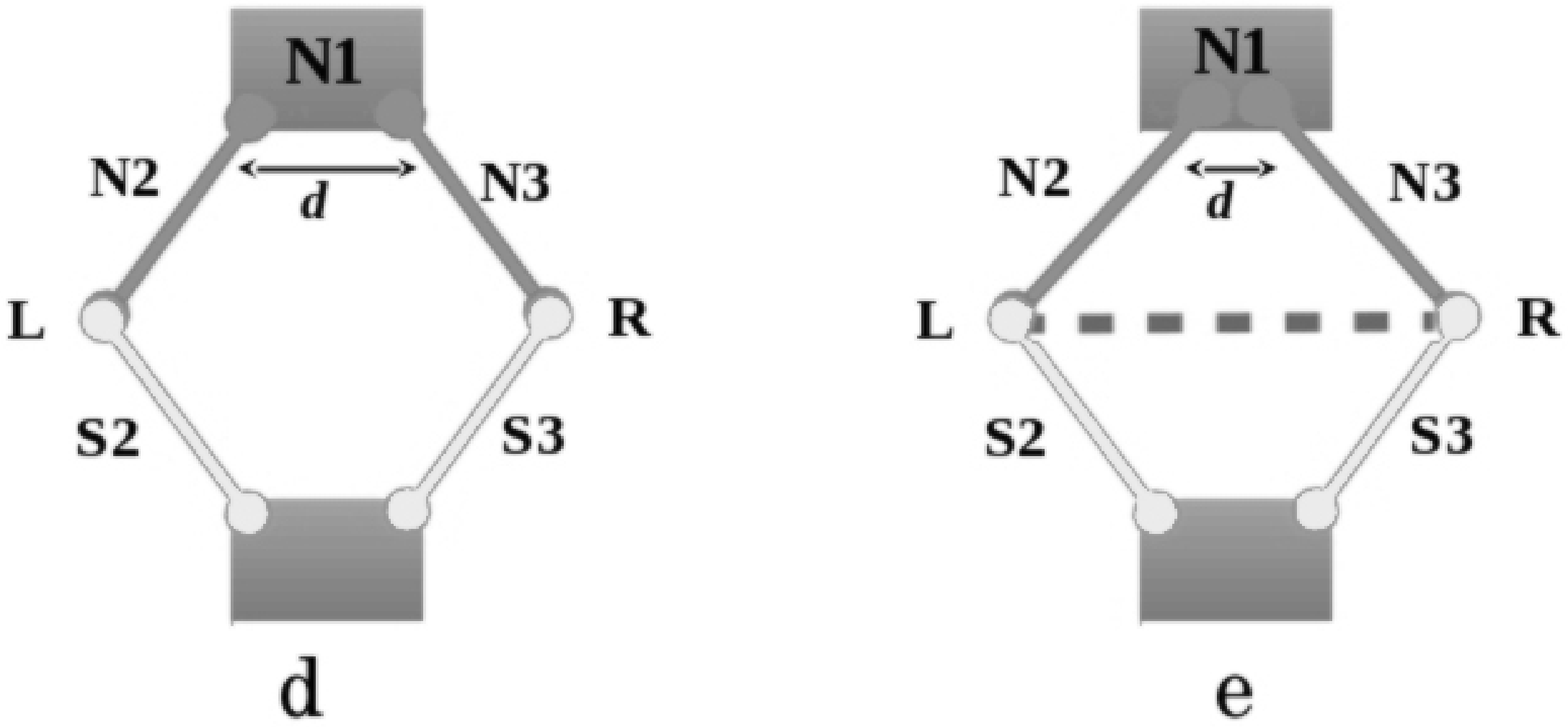}}
\vspace*{8pt}
\caption{The Braess paradox in a road network with two initial alternative routes (a), that modified by adding an extra link (b), and its quantum analog (c). Two last figures show the introduction of auxiliary AR and AL normal segments of vanishing lengths for the distance $d>L_\varphi$ (d) and $d<L_\varphi$ (e).}
\end{figure}

\section{Three-arm beam splitter}

The quantum transport in the proposed device includes specific type of a charge-transfer process at highly-transparent N/S interfaces by which a normal current in the N side is converted into a supercurrent in the S electrode, avoiding the forbidden single-particle transmission within the superconducting energy gap $\Delta $ (see Ref. 10 and references therein). To realize it, an incident from the N-side electron (e) with an energy $E$ around the Fermi energy $E_{\rm F}$ creates a Cooper pair in the superconductor with a second electron of the opposite spin which leaves a retroreflected hole (h) after that \cite{beloHbk}. The probability of such event is unity for $|\epsilon|<\Delta$ and rapidly goes to zero when $|\epsilon|\geq\Delta$, where $\epsilon=E-E_{\rm{F}}$. Reflection coefficients can be calculated from the boundary conditions for electron and hole wave functions from two sides of the interface. In this case, each Andreev quasielectron-into-quasihole transformation (and inverse) within the energy gap contributes an additional phase shift $\chi^{\rm{eh(he)}}(\epsilon)=-\arccos(\epsilon/\Delta)$ \cite{beloHbk}. As we show below, just the strong dependence on the energy permits to reveal the details of the charge transport across a hybrid system.

To make our calculations more comprehensible and as close as possible to the classical model by Braess shown in Figs. 1a and 1b we introduce two additional AR and AL normal paths the lengths of which are assumed to be negligibly small, see Fig. 1d. Now we have a pure normal three-arm part of the network formed by an ingoing electrode 1 and two AL and AR paths which are separated by a distance $d$. If $d$ strongly exceeds the electron decoherence length $L_\varphi$, the charge flow from the electrode 1 to two others is a trivial sum of currents in two independent channels (we name it a \emph{classical} case although in our case the transport in each channel is of quantum nature). When $d<L_\varphi$, all three electron waves are entangled and a new \emph{quantum} aspect of the problem emerges. The appearance of a quantum bond between the AL and AR paths is shown schematically by a dashed line in Fig. 1e and now the system looks like a classical one shown in Fig. 1b. As we argue below, the availability of this additional quantum link strongly modifies the transport characteristics of the device and generates reduction of the electron flux through the system which we name a \emph{quantum} Braess paradox.

Let us start with the discussion of a node with three converging normal 1, 2, and 3 leads and $d<L_\varphi$ (Fig. 1e) in perfect contact with each other and calculate the current flowing from an electrode 1 to the wires 2 and 3 which are grounded. We approximate the spectrum of electrons by parabolic bands, limit ourselves to a one-dimensional case and are dealing with wave functions carrying unit flux. In a non-superconducting metallic wire, the wave function of a quasiparticle excitation reads as $\psi_i(x)=\frac{m}{\hbar\sqrt{k_i}}exp(ik_ix)$, where $i$= 1,2,3; $m$ is the electron mass; $k_i=\sqrt{2m(E_{{\rm F}i}\pm\epsilon)/\hbar^2}$ is its wave vector in the $i$-th wire, $\epsilon $ is the quasiparticle energy measured with respect to the Fermi energy, the sign $\pm $ corresponds to electron and hole excitations, respectively. Next, we require the continuity of the wave functions and the conservation of the probability flux
\[
  \label{eq:0}
 j(x)=\frac{\hbar}{2 m i}\bigl( \psi^\star\frac{d \psi(x)}{dx} - \psi\frac{d \psi^\star(x)}{dx}\bigr) 
\]
at the discussed node at $x=0$: $\psi_0(0)=\psi_1(0)=\psi_3(0)=\psi_0={\rm {const}}$ and $j_1(0)-j_2(0)-j_3(0)=0$. Dividing both sides of the latter expression by the constant value $|\psi_0|^2$ that, of course, should be non-zero we obtain that
\begin{equation}
  \label{eq:1}
  \textmd{Im} \Bigl(\frac{1}{\psi_0}\frac{d \psi_1(x)}{dx}\bigl|_{x=0} - \frac{1}{\psi_0}\frac{d \psi_2(x)}{dx}\bigl|_{x=0} -\frac{1}{\psi_0}\frac{d \psi_3(x)}{dx}\bigl|_{x=0} \Bigr) = 0
\end{equation}

Thus, for an ideal contact of the three wires without any backscattering at the convergence point we get $ \frac{d \psi_1(x)}{dx}\bigl|_{x=0} - \frac{d \psi_2(x)}{dx}\bigl|_{x=0} - \frac{d \psi_3(x)}{dx}\bigl|_{x=0} = 0 $ and the probability amplitudes for the lead 1 read as
\begin{equation}
  \label{eq:2}
  t_{12}=\frac{2\sqrt{k_1k_2}}{k_1+k_2+k_3}; \quad t_{13}=\frac{2\sqrt{k_1k_3}}{k_1+k_2+k_3}; \quad r_{11}=\frac{k_1-k_2-k_3}{k_1+k_2+k_3}\;,
\end{equation}
where $t_{12}$ and $t_{13}$ are the transmission amplitudes for an electron transferring from the first to the second (third) lead and $r_{11}$ is the reflection amplitude for an electron returning back to the first electrode. Other scattering characteristics can be written in the same way.

The quantum contacts discussed below are a simple example of flow networks for which the well-known maximum-flow minimum-cut theorem is valid. It states that finding a maximal network flow is equivalent to finding a cut of minimum capacity that separates the source and the sink. In our case, it means that the maximal probability to transfer the system should be calculated for a transverse section of an incoming lead.  It follows from Eq. (2) that the probability of an electron to leave the wire 1 equals to $1-r^{\; 2}_{11}=4k_1(k_2+k_3)/(k_1+k_2+k_3)^2$. Imagine that the quantum link between the wires 2 and 3 is destroyed (Fig. 1d), then we have two independent channels with a total probability: $2k_1k_2/(k_1+k_2)^2+2k_1k_3/(k_1+k_3)^2$. There is a wide range of parameters when the latter value exceeds the previous one. For example, without any quantum link between the wires, Fig. 1d, the probability $D$ of an incoming electron to appear in the two outgoing leads equals to unity for $k_1=k_2=k_3$, while according to Eq. (2) $D$ is about 0.9 in the contrary case, Fig. 1e. It is just a manifestation of the \emph{quantum} Braess paradox when the introduction of a new transport channel leads to the reduction of the charge flow through the total system.

Now we shall return to the node shown in Fig. 1e and assume the presence of backscattering events at the convergence point $x$ = 0. If so, then from Eq. (1) it follows that
\begin{equation}
\label{eq:3}
\frac{d \psi_1(x)}{dx}\bigl|_{x=0} - \frac{d \psi_2(x)}{dx}\bigl|_{x=0} - \frac{d \psi_3(x)}{dx}\bigl|_{x=0} = K \psi_0
\end{equation}
with $K$, a real constant characterizing coupling of the wave functions at $x = 0$. Its origin can be simply explained for an N-I-N junction (I is an ultra-thin insulating interlayer which permits quantum-mechanical tunneling between the N terminals). In the latter case, the same relation with $K\equiv 2mH/\hbar^2$ arises for a $\delta$-functional potential $V(x) = H \delta(x)$ located at the I interface \cite{blond82}. If the number of wires connecting at one point is greater than two, the parameter $K$ is not so easily interpretable and it may be described as an ``effective potential barrier'' at $x = 0$. Then the related scattering amplitudes for normal leads (2) should be replaced by
\begin{equation}
  \label{eq:4}
  t_{12}=\frac{2\sqrt{k_1k_2}}{k_1+k_2+k_3+iK}; t_{13}=\frac{2\sqrt{k_1k_3}}{k_1+k_2+k_3+iK}; r_{11}=\frac{k_1-k_2-k_3-iK}{k_1+k_2+k_3+iK}\,.
\end{equation}
Together with other six amplitudes they form a scattering matrix
\[
\mathbf{S}^{\rm e}_{\rm N} = \left(
  \begin{array}{ccc}
    r^{\rm e}_{11} & t^{\rm e}_{12} & t^{\rm e}_{13} \\
    t^{\rm e}_{21} & r^{\rm e}_{22} & t^{\rm e}_{23} \\
    t^{\rm e}_{31} & t^{\rm e}_{32} & r^{\rm e}_{33} 
  \end{array}
\right)
\] 
for electrons and the complex conjugated matrix for holes in the normal part of the system. 

Our next aim is to propose a realistic experiment able to reveal partial suppression of the charge flow in the Y-shaped beam splitter after a quantum link between leads 2 and 3 is added. Below we argue that the difference between the two alternatives shown in Figs. 1d and 1e can be clearly identified by measuring the differential conductance spectrum of a normal metal-superconductor (or superconductors) device, i.e., the first derivative of the current $I$ with respect to the voltage $V$ $G(V) = dI(V)/dV$. 

Next we keep the basic notations for the transport problem similar to those used for solving it in the case of conventional NIS junctions \cite{belo99,belo03} and introduce the following matrices:
\[
\mathbf{T}_{{\rm e}({\rm h})}= \left(
    \begin{array}{c}
    t^{{\rm e}({\rm h})}_{12} \\
    t^{{\rm e}({\rm h})}_{13} 
  \end{array}
\right), \;
\mathbf{\widetilde{T}}_{{\rm e}({\rm h})}= \left(
    \begin{array}{cc}
      t^{{\rm e}({\rm h})}_{21} &  t^{{\rm e}({\rm h})}_{31} 
  \end{array}
\right), \;
\]
\[
\mathbf{R}_{{\rm e}({\rm h}} = \left(
  \begin{array}{cc}
    r^{{\rm e}({\rm h})}_{22} & r^{{\rm e}({\rm h})}_{23} \\
    r^{{\rm e}({\rm h})}_{32} & r^{{\rm e}({\rm h})}_{33}
  \end{array}
\right), \;
\mathbf{R}_{{\rm {eh}}({\rm {he}}}(\epsilon) = \left(
  \begin{array}{cc}
    \exp(i \chi(\epsilon)) & 0 \\
    0 & \exp(i \chi(\epsilon))
  \end{array}
\right).
\]
$\mathbf{R}_{{\rm {eh}}({\rm {he}})}(\epsilon)$ is an Andreev-scattering matrix obtained within the usual step-function approximation for a superconducting order parameter when self-consistency of its spatial variation is ignored (this approximation is valid for a conventional $s$-wave pairing). Similar to the approach described in Refs. 12 and 13, probability amplitudes can be found by summarizing all possible charge paths including Andreev transformations. For example, the probability amplitude for the retroreflected hole at the N-side of the normal metal-superconductor interface equals to
\[
{R}^{\rm {eh}}_{\rm N}(\epsilon)=\mathbf{\widetilde{T}}_{\rm h} \left(\mathbf{I} - \mathbf{R}_{{\rm {eh}}}(\epsilon) \mathbf{R}_{e} \mathbf{R}_{{\rm {he}}}(\epsilon) \mathbf{R}_{\rm h} \right)^{-1} \mathbf{R}_{{\rm {eh}}}(\epsilon) \mathbf{T}_{\rm e},
\]
compare with Eq. (3) in Ref. 13.

In our discussion for a hybrid superconducting device, we start again with the simplest situation of identical wavenumbers $k_1=k_2=k_3=k_0$ and two identical superconductors S$_2$ and S$_3$ with the energy gap $\Delta$ when, as was shown above, the Braess paradox is revealed itself. For a classical case $t_{12}=t_{13}=1$, $r_{23}=0$ and we get a step-like behavior with a pronounced feature at $V= \Delta / \epsilon$, well known for a standard NS point-contact \cite{blond82}, see Fig. 2a. In the quantum limit the three channels are identical and the charges may go from the second to the third wire with the same probability amplitude as from the first electrode to them $t_{12}=t_{13}=t_{23}= 2/3$. It results in finite reflection probability amplitudes $r_{12}= r_{13}=r_{23}= -1/3$ and, as a result, the $G(V)$ characteristic resembles that of a tunnel N-I-S junction with a low-height I barrier \cite{blond82}, see Fig. 2a. It is just a visual manifestation of a partial charge flow inhibition according to the quantum Braess paradox. When $K\neq0$ we observe a well pronounced peak at $V=\Delta /e$ in both limits and the difference between them is more quantitative (Fig. 2a).

\begin{figure}[h!]
\centerline{\includegraphics[width=.47\columnwidth]{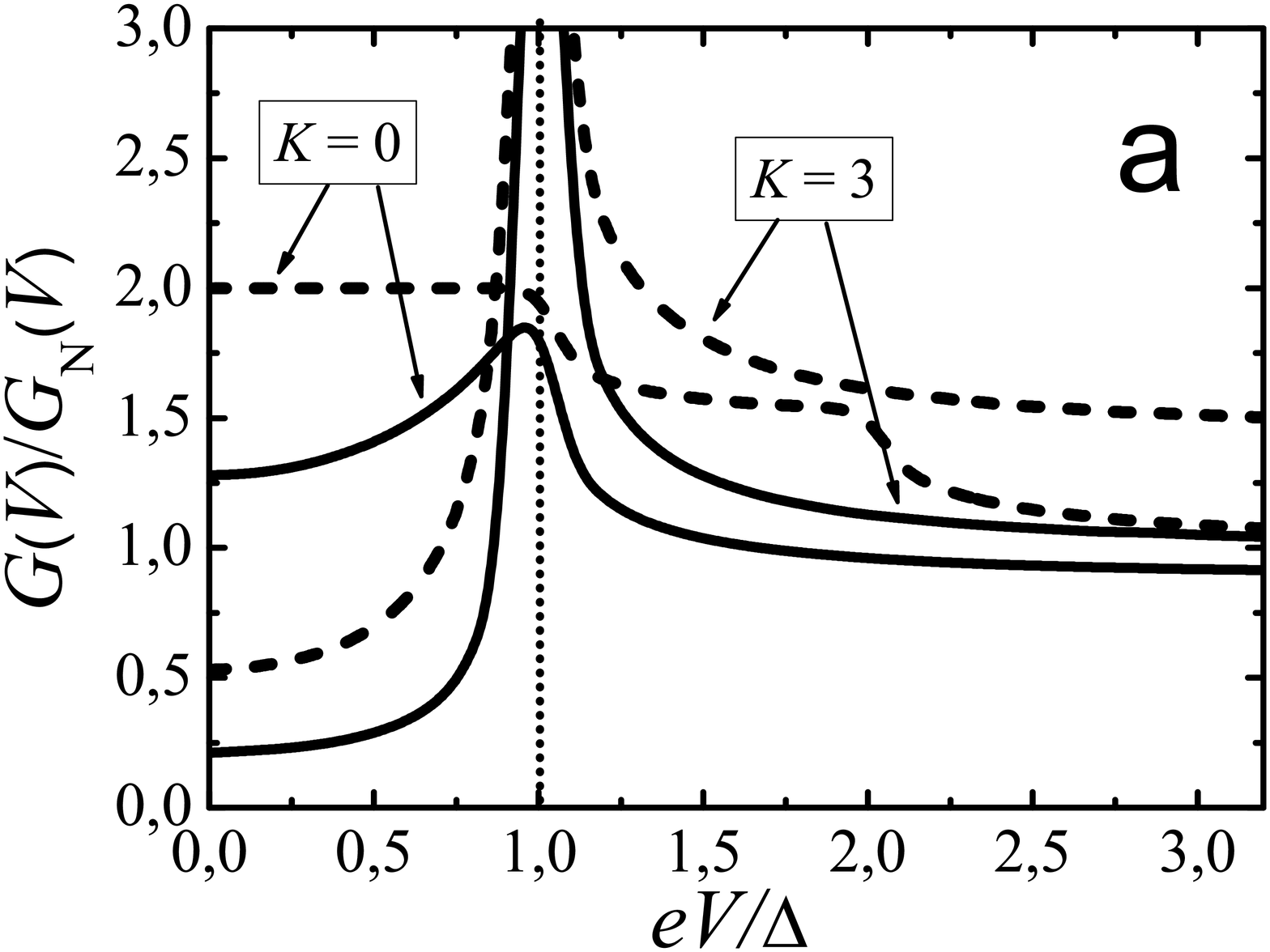}
\hspace*{8pt}
\includegraphics[width=.47\columnwidth]{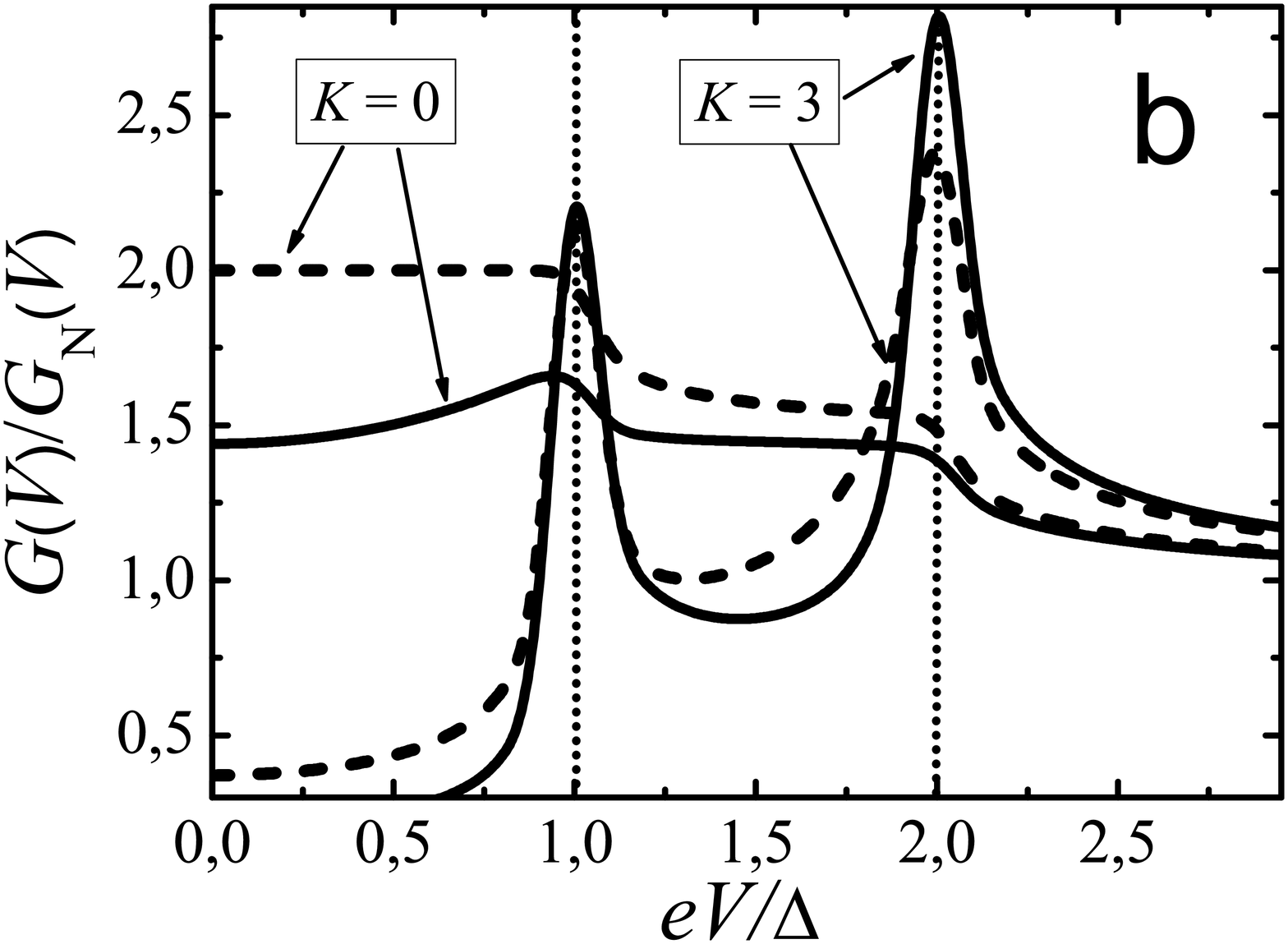}} 
\vspace*{8pt}
\caption{ Differential conductance spectra of a Y-shaped junction formed by  two identical superconducting wires with an ideal node, $K=0$ and an effective potential barrier, $K\neq0$. Classical results for $d>L_\varphi$ and quantum-limit ones for $d<L_\varphi$ are shown by  dashed and solid lines, respectively; $\Delta_2 =\Delta_3=\Delta$ (a) and  $\Delta_3$/$\Delta_2 = 2$, $\Delta_2 = \Delta$ (b). Dotted lines show positions of the energy gaps; $G_{\rm N}$ is the device conductance in the normal state.}
\end{figure}

Fig. 2b shows the difference between the two limits for identical wave vectors but different energy gaps $\Delta_2$ and $\Delta_3$ in the two superconducting leads. When the two S branches are isolated each from other, $t_{12}=t_{13}=1$ and $r_{23}= 0$, we get a well-known result \cite{brinkm02} for $G(V)$, Fig. 2b, that depends on the height of the potential barrier between N and S$_{2,3}$ electrodes. There should be two steps in the $G(V)$ curve at $V=\Delta_2/e$ and $V=\Delta_2/e$ when the barrier is absent (a point-like contact) and two peaks at the same voltage biases for strong scattering at the N/S$_{2,3}$ interfaces (a tunneling limit), Fig. 2b, dashed curves\cite{brinkm02}. These results are modified by introducing a coherent quantum link between the two S leads, Fig. 2b, solid curves and the difference between the classical and quantum limits can be easily detectable experimentally. As in Fig. 2a, the main changes occur for a point contact when $K=0$, whereas tunneling results shown in Fig. 2b exhibit no qualitative deviations. When $t_{23}\neq0$, the probability amplitude $t_{12}< 1$ and at voltages $V<\Delta_2/e$ we again have a high-transparent tunneling-like junction for which the standard BTK theory\cite{blond82} predicts a zero-bias conductance slightly above unity and a low-height peak at the gap value $V=\Delta_2/e$. The latter feature is not well pronounced since the related Andreev bound state is not formed inside the second channel due to the charge penetration into the third one. At $V>\Delta_2/e$ new processes arise. Andreev scattering in the second wire is quickly suppressed down \cite{blond82} and the contribution of the second channel into the total current strongly decreases. At the same time, another process diminishing the Andreev-scattering effect in the third channel starts to manifest itself, namely, only a part of holes reflected back is going to the first channel while another part goes to the second channel and disappears leaving it at $V>\Delta_2/e$. At these bias voltages, new quasiparticle round-trips are emerging, for example, electron-into-hole scattering in the third channel followed by the hole-into-electron transformation in the second channel and electron transferring back to the third one. 

The asymmetry in the normal-state characteristics of the two S channels strongly modifies the shape of the $G(V)$ characteristic. In Fig. 3a we demonstrate how the conductance spectrum shown in Fig. 2b is changed for $k_2\neq k_3$. It can be seen that the wavenumber mismatch growth increases backscattering amplitudes $r_{21}$ or $r_{31}$ \cite{blond16}, enhances the amplitude of an Andreev bound state formed at the N/S interface \cite{zhitl16} and, as a result, enlarges the peak at $V_2 = \Delta_2/e$ or $V_3 = \Delta_3/e$, respectively, see Fig. 3a.
\begin{figure}[h!]
\centerline{\includegraphics[width=.47\columnwidth]{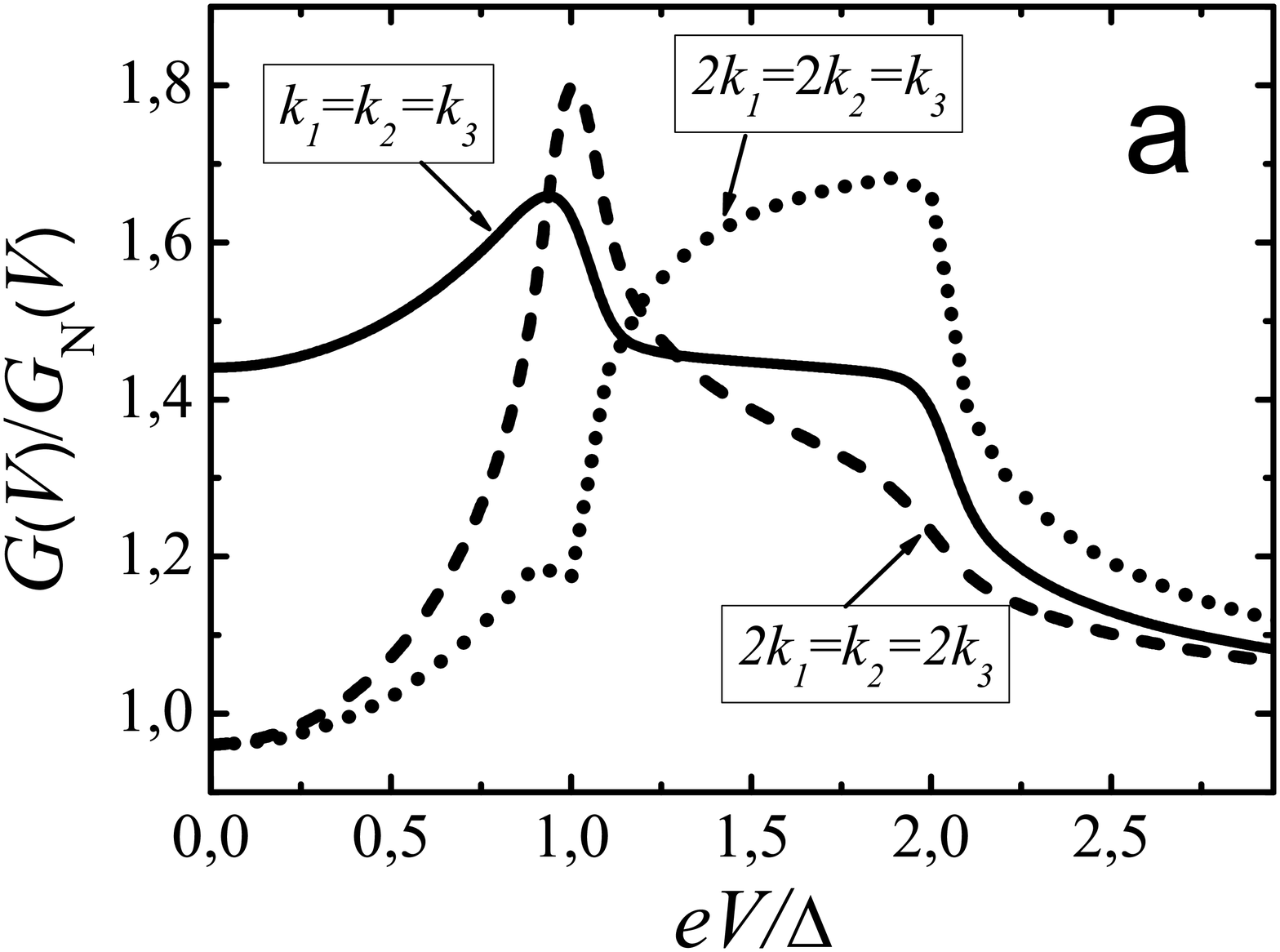}
\hspace*{8pt}
\includegraphics[width=.47\columnwidth]{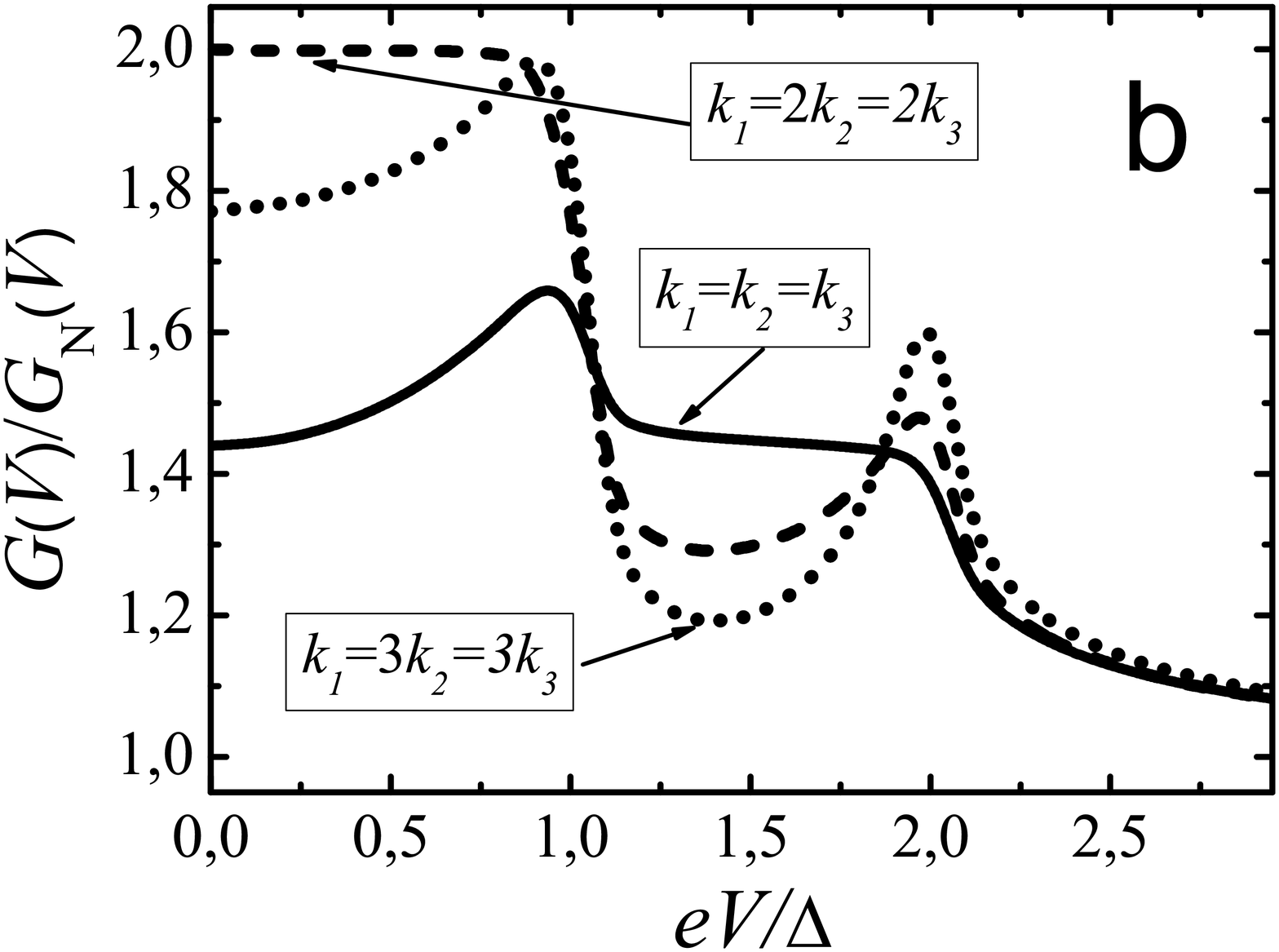}} 
\vspace*{8pt}
\caption{Effect of wavenumbers $k_i \; (i=1,2,3)$ on the quantum-limit conductance spectra of a Y-shaped junction for two different superconductors with $\Delta_3/\Delta_2 = 2$, $\Delta_2 =\Delta$ and $K=0$; the values of $k_i$ are indicated in the figures; the dashed lines show positions of the energy gaps; $G_{\rm N}$ is the normal-state device conductance.}
\end{figure}

Fig. 3b shows that some results for a three-arm device can be unexpectable due to an accident coincidence of values. For example, for wavenumbers $k_1= 2k_2= 2k_3$ the reflection amplitude and the conductance spectrum below $\Delta_2$/$e$ looks like that for a classical case shown by a dashed curve in Fig. 2b, whereas further increase in $k_1$ with respect to $k_2$ and $k_{3}$ leads, as it should be, to tunneling-like curves, see the $G$($V$) characteristic for wavenumbers $k_1= 3k_2= 3k_3$ in Fig. 3b.

Resuming, we have shown that phase-coherent quantum transport across a Y-shaped metallic junction reveals the main manifestations of the Braess paradox, namely, reduction of the complex network transparency after addition of a new link between two outgoing leads. It follows that the backscattering in the three-arm structure takes place even without any scattering at the node and for identical leads. As a result, in a quantum situation the current through the system can be significantly modified comparing to the classical case of $d >L_{\varphi}$, when the total transmission probability is a sum of partial contributions. To reveal this effect experimentally in a normal-state fork is a difficult task and we have proposed an experiment with two superconducting wires as the outgoing channels. It was shown that in this case the shape of the differential conductance-versus-voltage curves is visually modified when $D$ decreases due to the introduction of a new quantum bond and it can be considered as a manifestation of a Braess-like paradox, originating precisely from quantum effects.

It is evident that our discussion concerning a Y-shaped structure characterized by the scattering matrix $\mathbf{S}^{\rm e}_{\rm N}$ is equivalent to that for a problem of the charge current into a two-band superconductor. Thus, the results presented above can be applied for explaining related experimental data when they differ from the analysis based on two separate and independent charge groups \cite{brinkm02}.

At last, we believe that this study paves the way towards experiments establishing the presence and amount of quantum entanglement in superconducting interconnects as well as to estimation of a decoherence length in such quantum networks. Recent progress in the fabrication of ultra-thin metallic wires with favorable characteristics permits to focus the analysis on the ballistic limit treating it exactly and comprehensibly.

\section{Acknowledgments}

 E. Zhitlukhina is sincerely grateful to the German Academic Exchange Service (DAAD) for the support of her stay at the Friedrich-Schiller University Jena and the possibility to conduct this research in a friendly international team. M. Belogolovskii is thankful to the support of the Fundamental Research Programme funded by the MES of Ukraine (Project No. 0117U002360).

\end{document}